\documentclass[%
 reprint,
groupedaddress,
 amsmath,amssymb,
 aps,
 longbibliography,
]{revtex4-1}

\providecommand\bnabla{\boldsymbol{\nabla}}
\providecommand\bkappa{\boldsymbol{\kappa}}

\newcommand{\ensbl}[1]{\left\langle #1 \right\rangle}

\newcommand{\vpar}{\ensuremath{v_{\parallel}}}
\newcommand{\vperp}{\ensuremath{v_{\perp}}}
\newcommand{\xpar}{\ensuremath{x_{\parallel}}}

\def \al {\mbox{$\alpha$}}

\def \Lpar {\mbox{$L_{\parallel}$}}
\def \vth {\mbox{$v_{\mathrm{T}}$}}
\def \vp {\mbox{$\varphi$}}

\def \l {\mbox{$\ell$}}
\def \lp {\mbox{$\ell^{\prime}$}}

\renewcommand \o {\mbox{$\omega$}}

\def \od {\mbox{$\omega_d$}}

\def \Reff {\mbox{$R_{\mathrm{eff}}$}}

\usepackage{graphicx}
\usepackage{dcolumn}
\usepackage{bm}
\usepackage{amsmath}
\usepackage{color}

\begin{document}

\title{Coarse-grained gyrokinetics for the critical ion temperature gradient in stellarators}

\author{G. T. Roberg-Clark}
\email{gar@ipp.mpg.de}
\author{G. G. Plunk}
\author{P. Xanthopoulos}
\affiliation{Max-Planck-Institut F\"ur Plasmaphysik, D-17491, Greifswald, Germany}


\date{\today}

\begin{abstract}

We present a modified gyrokinetic theory to predict the critical gradient that determines the linear onset of the ion temperature gradient (ITG) mode in stellarator plasmas. A coarse-graining technique is applied to the drift curvature, entering the standard gyrokinetic equations, around local minima. Thanks to its simplicity, this novel formalism yields an estimate for the critical gradient with a computational cost low enough for application to stellarator optimization.  On comparing against a gyrokinetic solver, our results show good agreement for an assortment of stellarator designs. Insight gained here into the physics of the onset of the ITG driven instability enables us to devise a compact configuration, similar to the Wendelstein 7-X device, but with almost twice the ITG linear critical gradient, an improved nonlinear critical gradient, and reduced ITG mode transport above the nonlinear critical gradient.

\end{abstract}

\date{\today}

\maketitle


\textit{Introduction.}--  The excitation of the ion temperature gradient (ITG) mode in magnetically confined fusion devices leads to turbulence that is responsible for energy losses, which deteriorate plasma confinement. For instance, it has been argued that, during operation of the Wendelstein 7-X (W7-X) stellarator in electron heating scenarios, the ITG mode leads to so-called ``ion temperature clamping"  \citep{Beurskens2021a}, thus preventing the heating of ions in the plasma core above 2 keV. In order to lessen the negative effects of the ITG mode, a possible strategy to follow involves the manipulation of the density and electron profiles, as implemented in W7-X using pellet injections \citep{Bozhenkov2020a}. 

The problem might also be faced at the stage of stellarator design, by shaping the magnetic field to reduce losses from core micro-turbulence. Such strategies can already be applied to target electron-temperature gradient and trapped-electron mode turbulence, the former suppressed by multiple field periods \citep{Plunk2014a} and the latter by optimization for the maximum-J property \citep{Proll2012}. Magnetic field shaping is the motivation behind turbulence optimization studies, which target the rate of the ITG transport increase (``stiffness") as a function of the ion temperature gradient \citep{Mynick2010a,Xanthopoulos2014a,Hegna2018,Nunami2013}. 

To complement this effort, one can identify another strategy: to target the threshold or ``critical gradient'' of the mode itself.  This should be an effective strategy for such cases where transport is stiff, as in the aforementioned ion clamping scenario of W7-X, such that the rapid onset of turbulence relaxes the temperature gradient back to its threshold \citep{Biglari1989a}, essentially determining the temperature profile allowed in the plasma \citep{Baumgaertel2013}.  With the actual onset of turbulence expected to differ by a positive increment from the linear ITG threshold \citep{Dimits2000}, the linear critical gradient can be used as a lower-bound estimate of the nonlinear critical gradient, indeed, one that is much simpler to model and compute. 

Sustained efforts over the past four decades \citep{Terry1982,Hahm1989a,Dominguez1989a,Romanelli1989,Biglari1989a,Kadomtsev1995a, Zocco2018} have yielded an understanding of the linear onset of the instability in toroidal geometry, forming the basis of tokamak models \cite{Jenko2001a, Zocco2018}. In this Letter, we propose a model of the toroidal ITG critical gradient in generic stellarator geometry.  This is achieved by spatially coarse-graining the gyrokinetic equations under the assumption of scale separation, which, as inferred from our results, appears to be valid in a large class of stellarator optimization lines.  We then apply the model findings to a novel configuration, generated by modifying the large scale properties of the W7-X stellarator, that demonstrates a significant increase in the critical gradient as well as reduced turbulent transport above the critical gradient.

\textit{Definitions.}--  Following \citep{Plunk2014a}, we use the standard gyrokinetic system of equations \citep{Brizard2007} to describe electrostatic fluctuations destabilized along a thin flux tube tracing a magnetic field line. The ballooning transform \citep{Dewar1983a} and twisted slicing representation \citep{Roberts1965} are used to separate out the fast perpendicular (to the magnetic field) scale from the slow parallel scale. The magnetic field representation in field following (Clebsch) representation reads, $\mathbf{B}=\bnabla \psi \times \bnabla \al$, where $\psi$ is a flux surface label and $\alpha$ labels the magnetic field line on the surface. The perpendicular wave vector is then expressed as $\mathbf{k_{\perp}} = k_{\alpha} \bnabla \alpha + k_{\psi} \bnabla \psi$, where $k_{\alpha}$ and $k_{\psi}$ are constants, so the variation of $\mathbf{k_{\perp}}(l)$ stems from that of the geometric quantities $\bnabla \alpha$ and $\bnabla \psi$, with $\l$ the field-line-following (arc length) coordinate.

We assume Boltzmann-distributed (adiabatic) electrons, thus solving for the perturbed ion distribution $g_{i}(\vpar,\vperp,\l,t)$, defined to be the non-adiabatic part of $\delta f_{i}$ ($\delta f_{i}=f_{i}-f_{i0})$ with $f_{i}$ the ion distribution function and $f_{i0}$ a Maxwellian. The electrostatic potential is $\phi(\mathbf{\l})$, and $\vpar$ and $\vperp$ are the particle velocities parallel and perpendicular to the magnetic field, respectively. The integral equation for $g$ \citep{Plunk2014a} reads

\begin{equation}\label{eqn:g-int}
    g(\ell)=-i \sigma (\omega-\omega_{*}^{T})f_{0}\int^{\ell}_{-\sigma \infty}d \lp \frac{J^{'}_{0}}{|\vpar^{'}|}\vp(\lp)\text{exp}(i\sigma M(\ell,\lp))
\end{equation}
with
\begin{equation}\label{eqn:M}
M(\ell^{\prime}, \ell) = \int_{\ell^{\prime}}^{\ell} \frac{\o - \od(\ell^{\prime\prime})}{\vth \xpar } d\ell^{\prime\prime}
\end{equation}
where $\sigma=\text{sgn}(\vpar)$, $\omega$ is the mode frequency, $\omega_{*}^{T} = (Tk_{\alpha}/q)d\ln T/d\psi \left(v^2/\vth^2 - 3/2\right)$ is the diamagnetic frequency, and $J^{\prime}_{0} = J_{0}(k_{\perp}(\lp)v_{\perp}/\Omega(\lp))$ is the Bessel function of zeroth order. The thermal velocity is $\vth = \sqrt{2T/m}$, the thermal ion Larmor radius is $\rho = \vth/(\Omega\sqrt{2})$, $n$ and $T$ are the background ion density and temperature, $q$ is the ion charge, $\varphi = q\phi/T$ is the normalized electrostatic potential, $\Omega=q B/m$ is the cyclotron frequency, with $B=|\mathbf{B}|$, and $\vpar^{\prime}=\vpar(\lp)$, where  $\lp$ is a dummy variable of integration along the field line. The magnetic drift frequency in the low $\beta$ approximation becomes $\od = (1/\Omega)(\mathbf{k_{\perp}} \cdot \mathbf{b} \times \bkappa) (\vpar^2 + \vperp^2/2)$, with $\bkappa = \mathbf{b} \cdot\bnabla\ \mathbf{b}$ and $\mathbf{b}=\mathbf{B}/B$. We select the magnetic field line with $\alpha=0$, thus setting $k_{\psi}=0$. Finally, we define a radial coordinate $r=a\sqrt{\psi/\psi_{edge}}$, with $a$ the minor radius corresponding to the flux surface at the edge, and $\psi_{edge}$ the toroidal flux at that location.  As usual, the temperature gradient scale length is measured relative to the minor radius, $a/L_{T}=-(a/T)dT/dr$. To study the most unstable ITG mode conditions, we neglect certain stabilizing factors such as the density gradient \citep{Proll2022a,Jorge2021b} and plasma beta (electromagnetic effects) \citep{Pueschel2008a}.

Note that Eq.~(\ref{eqn:g-int}) is equivalent to the ion gyrokinetic equation under the assumption of ballooning boundary conditions along the field line \citep{Connor1980}.  The system is completed by quasineutrality,

\begin{equation}
\int d^3{\bf v} J_{0} g = n(1 + \tau) \varphi\label{eqn:qn},
\end{equation}
with $\tau=q_{e}T_{i}/(q_{i}T_{e})$, $T_{e}$ the electron temperature, and $q_{e}$ the electron charge.

\textit{Scale separation and coarse-grained theory.}--  Spatial variation due to magnetic geometry appears in two places in the above equations, namely $k_\perp$ in the argument of $J_0$, and the magnetic drift frequency $\od$, from which a purely geometric quantity can be factored out, $\od \propto K_d = a^2{\mathbf\nabla}\alpha \cdot \mathbf{b} \times \bkappa$, here called the ``drift curvature''. We exploit the fact that the drift curvature contains large-scale variation ({\em e.g.,} due to global magnetic shear, toroidal variation of the normal curvature) as well as short-scale variation ({\em e.g.,} helical ripple, local shear).  We denote these macro- and microscales as $L$ and $\ell_r \ll L$, respectively.  The macroscale is assumed to be comparable to the linear parallel correlation length, $\Lpar = \vth/\omega$, and the global magnetic shear length $L_{s}=|q R/\hat{s}|$, where $R$ is the major radius, $\hat{s} = (r/q)(dq/dr)$, and $q(r)$ is the safety factor.

We introduce a coarse-graining operator $\langle \cdot \rangle$ along the parallel direction, which can be regarded as an average over a spatial interval that is intermediate in size to $L$ and $\ell_r$.  We can, therefore, decompose any spatial function as  $\varphi = \ensbl{\varphi} + \delta \varphi$, so that $\ensbl{\delta\varphi} = 0$ by definition.

The averaging procedure arises in a straightforward manner in Eq.~(\ref{eqn:g-int}):  We are here interested in localized ITG modes, {\em i.e.}, modes whose amplitudes $|\varphi|$ peak within a confining well, defined roughly by a region of drift curvature of destabilizing sign ($\omega_{*}^{T}\od > 1$), and decay over a distance $\sim L$.  Fixing $\ell$ to be in such a region, we see that the behavior of the gyrokinetic system, Eqs. (\ref{eqn:g-int})-(\ref{eqn:qn}), is such that $M < 1$ is satisfied within the confining well (for $|\ell - \ell^\prime| \lesssim L$) and $M>1$ outside, so that the contribution to $\varphi$ is negligible under velocity integration in Eq.~(\ref{eqn:qn}) (see also the discussion in \citet{Roberg-Clark2021a}). Furthermore, we assume $J_{0} \simeq 1$ within this region as its argument is small there. Thus, for $|\ell - \ell^\prime| < L$, the integral in Eq.~(\ref{eqn:M}) acts to average over small scale variation in $\od$, except for an asymptotically small interval in $\ell^\prime$, where $|\ell - \ell^\prime| \sim \ell_r$, allowing us to take $M \approx \ensbl{M}+ \delta M$, with $\delta M$ a small correction.  Noting that $M \sim 1$, we write $\exp(i \sigma M) \approx \exp(i \sigma \ensbl{M})(1 + i \sigma \delta M)$, and thus find that we can express $g = \ensbl{g} + \delta g$, where $\delta g \ll \ensbl{g}$.

Applying the coarse-graining operator to the integral equation (\ref{eqn:g-int}), introduces the averaged quantities $\ensbl{g}$ , $\ensbl{\varphi}$, and $\ensbl{\od}$ which vary on the scale $L$. The same is true when applied to the quasineutrality condition, Eq.~(\ref{eqn:qn}), and the substitution of Eq.~(\ref{eqn:g-int}) into Eq.~(\ref{eqn:qn}) produces a single integral equation, identical in form to full linear gyrokinetic theory, except in coarse-grained variables. We conclude that solutions of the linear gyrokinetic equation using a coarse-grained drift curvature should give comparable answers to that of the original geometry, with errors on the order $\mathcal{O}(\l_{r}/L)$.

\begin{figure}
    \centering
    \includegraphics[scale=.65]{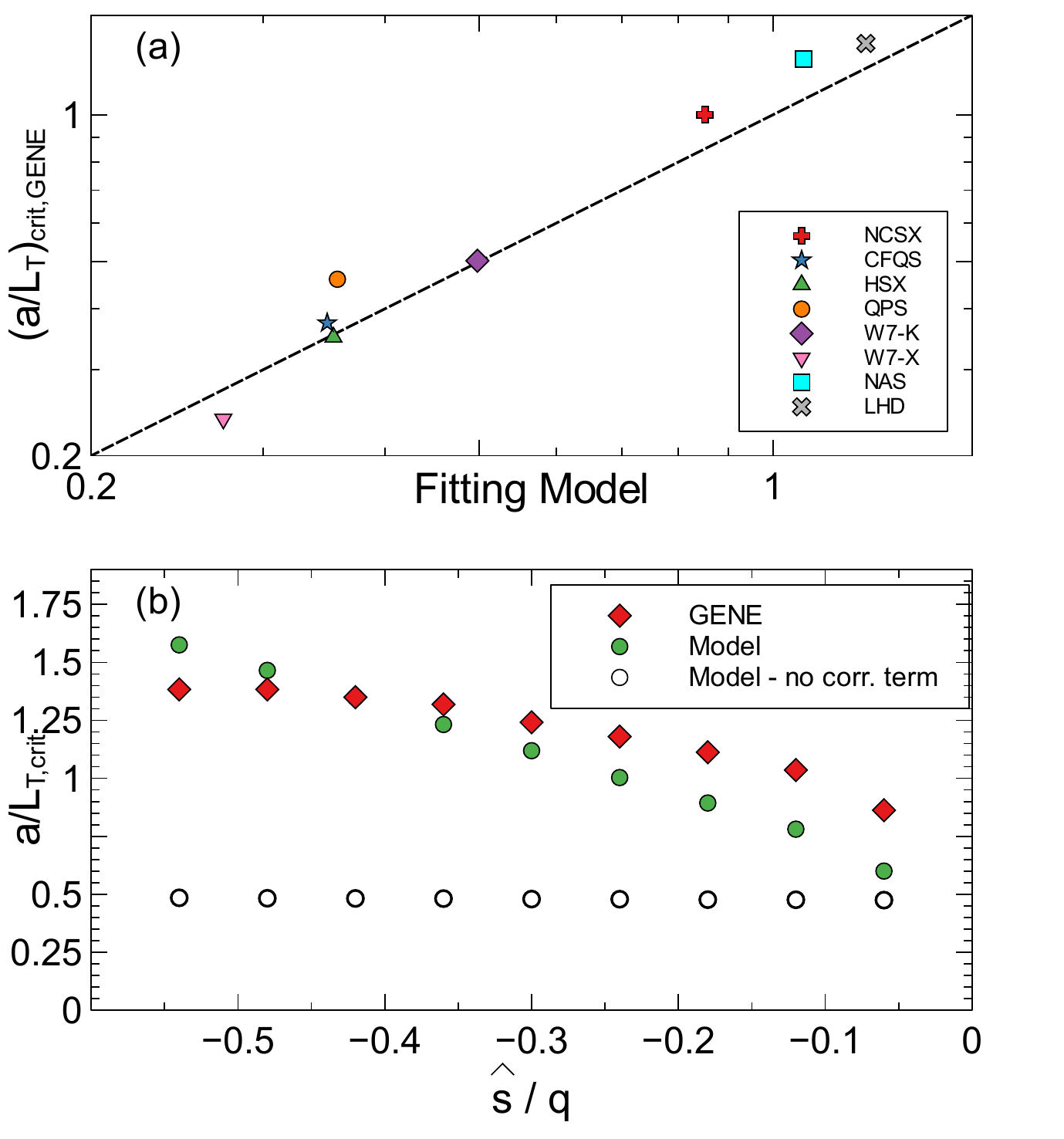}
    \caption{Simulation of critical gradients versus model predictions on a log-log scale for flux tubes taken from several stellarator geometries. (a) GENE critical gradients versus the model equation (\ref{eqn:critgrad}) prediction for a series of stellarator shapes, indicated by various shapes and colors. The theoretical line $a/L_{T,\text{crit}}=2.66 (a/\Reff+8.00 a/\Lpar)$ [Eqn. (\ref{eqn:critgrad})] is overlaid. (b) Predicted critical gradients from the model, including (green circles) and excluding (empty circles) the term $\propto \Lpar^{-1}$, as compared with critical gradients obtained from GENE simulations (red diamonds), using a series of flux tubes taken from a negative shear circular tokamak. Each value of $\hat{s}/q$ on the horizontal axis corresponds to a distinct flux tube.}
    \label{fig:critgrad}
\end{figure}
\textit{Critical gradient model.}-- The coarse-grained theory is now applied to calculate the ITG critical gradient. Our starting point is the drift resonance condition, $\omega_{*}^{T} \sim \od$ \citep{Biglari1989a}, which describes a balance between the drive from the temperature gradient $\omega_{*}^{T} \sim k_{\alpha}\rho v_{T}/L_{T}$ and resonant damping from drift curvature $\od \sim k_{\alpha}\rho v_{T}/\Reff$, where $\Reff$ is an effective radius of curvature. It expresses the simple fact that increasing the magnitude of drift curvature will raise the threshold gradient for instability \cite{Baumgaertel2013}. In the case of a circular-cross-section tokamak with large aspect ratio, the radius of curvature can be replaced by the major radius $R$, as was used by Jenko et al. \cite{Jenko2001a}. For ITG modes, ignoring shear and finite aspect ratio effects, the Jenko formula reads $R/L_{T,\text{crit}}=2.66$, assuming equal ion and electron temperatures and charges. For general geometry, such a formula cannot be expected to be accurate, but we argue that, in cases where localized modes set the threshold, and our coarse-grained theory applies, the replacement of $R$ with a modeled $\Reff$ should yield a good estimate.

\begin{figure}
    \centering
    \includegraphics[scale=.55]{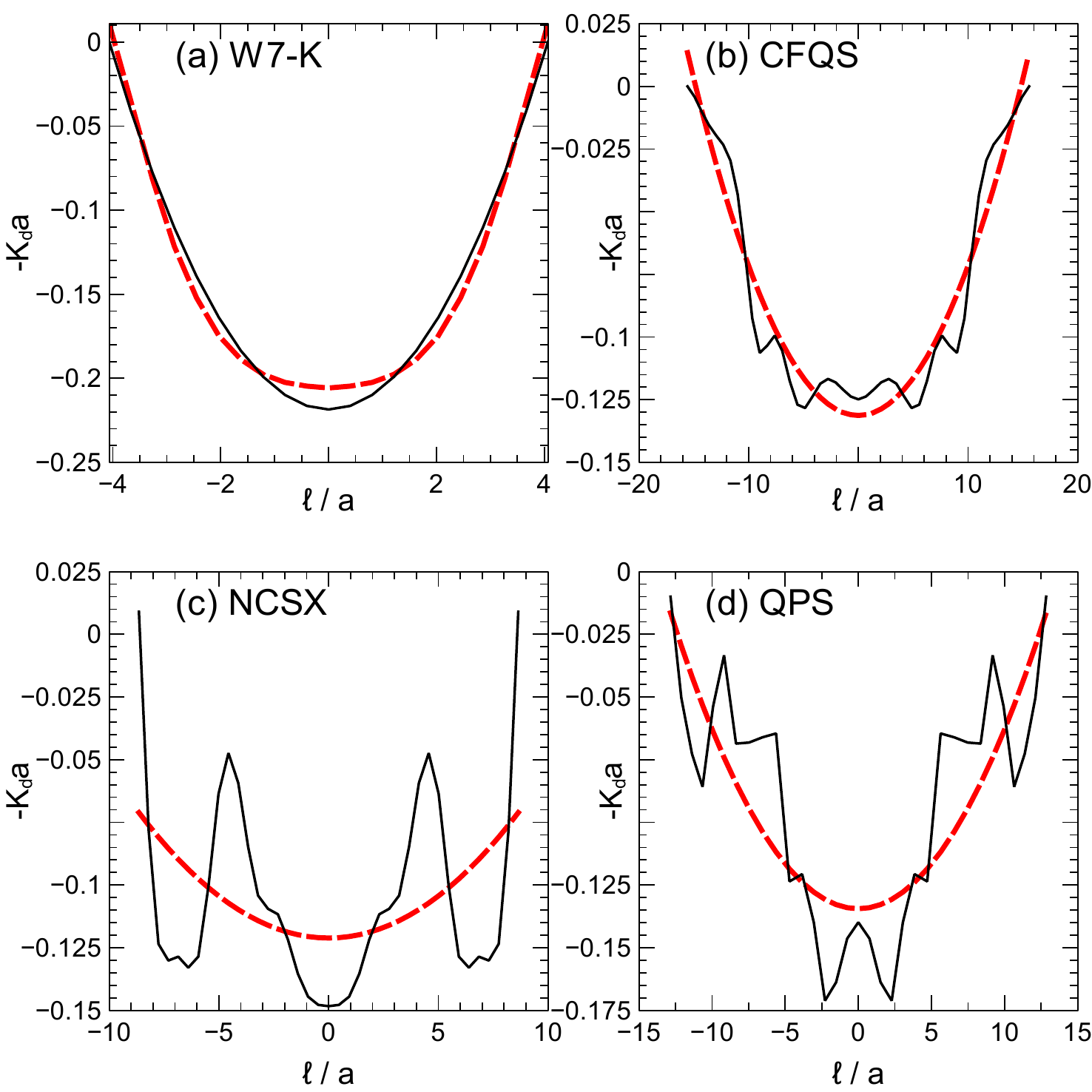}
    \caption{Fitting quadratic curves (dashed red curves) to the drift curvature profile $-K_{d} a$ versus $\l/a$ (solid black curves) used in the coarse-grained model for selected stellarator geometries.}
    \label{fig:fitting}
\end{figure}

To calculate $\Reff$, we take a drift curvature profile along a thin magnetic flux tube, which extends a number of poloidal turns needed to adequately sample a given magnetic field geometry, and find each point where the sign of the drift curvature $K_d$ changes.  This defines a series of local ``drift wells'' along the flux tube. Coarse-graining in $\l$ is implemented through a least-squares quadratic fit of $K_d(\ell)$, of the form $K_{d,n}(\ell)=\Reff_{,n}^{-1}(1-(\l-\l_{c})^{2}/\l_{\text{n}}^2)$, for the drift curvature profile within the $n$th well, where $\l_{c}=(\l_{\text{min}}+\l_{\text{max}})/2$ is the point at the center of the drift well and $\Reff_{,n}$ and $\l_{\text{n}}$ are free fitting parameters (see Fig.~\ref{fig:fitting}). We exclude inverted wells, {\em i.e.}, those where the sign of the drift curvature corresponds to stable ``good'' curvature wells. Thus, the effective radius of curvature corresponds to the peak value of the coarse-grained ``bad'' drift curvature. 

We find that an estimate for the critical gradient based purely on the above calculation of $\Reff$ accurately describes most stellarator configurations, except for cases where the ITG mode is significantly stabilized by a finite parallel correlation length  \citep{Jenko2000a,Hahm1989a}.  A typical estimate of this length is  the distance between regions of good and bad curvature, {\em i.e.}, the parallel connection length, but this does not account for necessity of global magnetic shear, which causes secular growth of drift curvature along the field line.  We thus propose an alternative definition $\Lpar^{-1} \sim \Reff^{-1}\Theta(p_{1}) \Theta(\Reff/R_{\text{good}}-1)/(\Reff/R_{\text{good}}+1) \times 0.5(1+\tanh[{20\:p_{2}}])$, where $L_{w}=\l_{\text{max}}-\l_{\text{min}}$ and $R_{\text{good}}$ is the scale obtained by the same fitting method as for $\Reff$, but applied to the neighboring region of good curvature. The quantity $p_{1}=\Reff/L_{w}-0.20$ is calculated to check for a sufficiently narrow drift well ($\Theta(x)=x H(x)$ with $H$ a Heaviside step function) and $p_{2}=\Reff/L_{s}-0.15$ must be positive so that the mode cannot extend beyond the neighboring good curvature well, which occurs when the growth of the drift curvature is only transient. The term has the expected physical behavior $1/\Lpar \rightarrow 0$ as $\Reff/R_{\text{good}} \rightarrow 1$ (periodic limit of Floquet modes \cite{Bhattacharjee1983, Zocco2018}) and $1/\Lpar \sim 1/L_w$ as $\Reff/R_{\text{good}} \rightarrow 0$ (perfect confinement of the mode inside the region of good curvature \cite{Plunk2014a}). Taking the above considerations into account, the final model with the correction is given by
\begin{equation}\label{eqn:critgrad}
 \frac{a}{L_{T,crit}}=2.66\left(\frac{a}{\Reff} + 8.00 \frac{a}{\Lpar}\right).   
\end{equation}
where $\Reff$ and $\Lpar$ are calculated from the drift well which minimizes $a/L_{T}$. The constant is obtained for the parallel term by rough calibration with a series of flux tubes taken from a negative shear circular tokamak with aspect ratio $3.87$, in which $q=2.22$ is held fixed and $\hat{s}$ is varied from $-0.20$ to $-1.20$ by scanning over the plasma radius, see Fig. \ref{fig:critgrad}(b).

The method described above, in accordance with the electrostatic limit of gyrokinetics, is applied to vacuum flux-tube geometries generated for an assortment of stellarator designs, including HSX \citep{Talmadge2008a}, CFQS \citep{Shimizu2022a}, NCSX \citep{Zarnstorff2001a}, LHD \citep{Takeiri2017a}, QPS \citep{Nelson2003a}, the high-mirror configuration of Wendelstein 7-X \citep{Pedersen2016a}, a quasi-axisymmetric stellarator NAS \citep{Plunk2018}, and a modified high-mirror W7-X (KJM) case that we call W7-K, described in the next section. In Fig. \ref{fig:critgrad}(a) we plot the estimated critical gradients from the formula (\ref{eqn:critgrad}) along the horizontal axis versus the critical gradient extracted from linear flux-tube gyrokinetic calculations \citep{Beer1995a} using the GENE code \citep{Jenko2001a}. Flux tubes are constructed \citep{Xanthopoulos2009} with 8 poloidal turns except for the case of NCSX (4 turns) to obtain convergence of the critical gradient, as short flux tubes tend to artificially destabilize modes which extend far along the field line \citep{Faber2018}, especially in low-shear geometries. The chosen flux tube geometries are centered at the outboard midplane of the devices ($\alpha=0$, zero toroidal angle), and radially located at $r^{2}/a^{2}=0.5$. The critical gradient is obtained from the GENE simulations by tracking the most unstable mode and extrapolating its growth rate to $\gamma=0$ with two data points very close to marginality, {\em i.e.} $\gamma \simeq 10^{-3} \sqrt{T/m}/a$ or smaller for each data point. The marginally unstable modes tend to have values of the poloidal wavenumber that fall in the range $k_{y}\rho \simeq 0.2$ to $0.4$. The LHD heliotron has the highest critical gradient of all cases considered, as a result of significant drift curvature $a/\Reff \simeq 0.45$. Interestingly, its highly favorable comparison with W7-X (a factor of $\sim 7$ greater) is in line with the difference in nonlinear heat fluxes obtained via gyrokinetic simulations \citep{Warmer2021a}.

Several cases of the fitted drift curvature are depicted in Fig. \ref{fig:fitting} to demonstrate how the method behaves for four different stellarator geometries. The case of CFQS, Fig. \ref{fig:fitting}(b), is an ideal application of the coarse-grained approach, showing that small-scale ripples in the drift curvature can be removed and revealing the underlying parabolic structure of the well. Other cases [Fig. \ref{fig:fitting} (c)-(d)] show that the fit can strongly deviate from the exact profile -- indeed the fluctuations need not be small according to our theory.

\begin{figure}
    \centering
    \includegraphics[scale=.25]{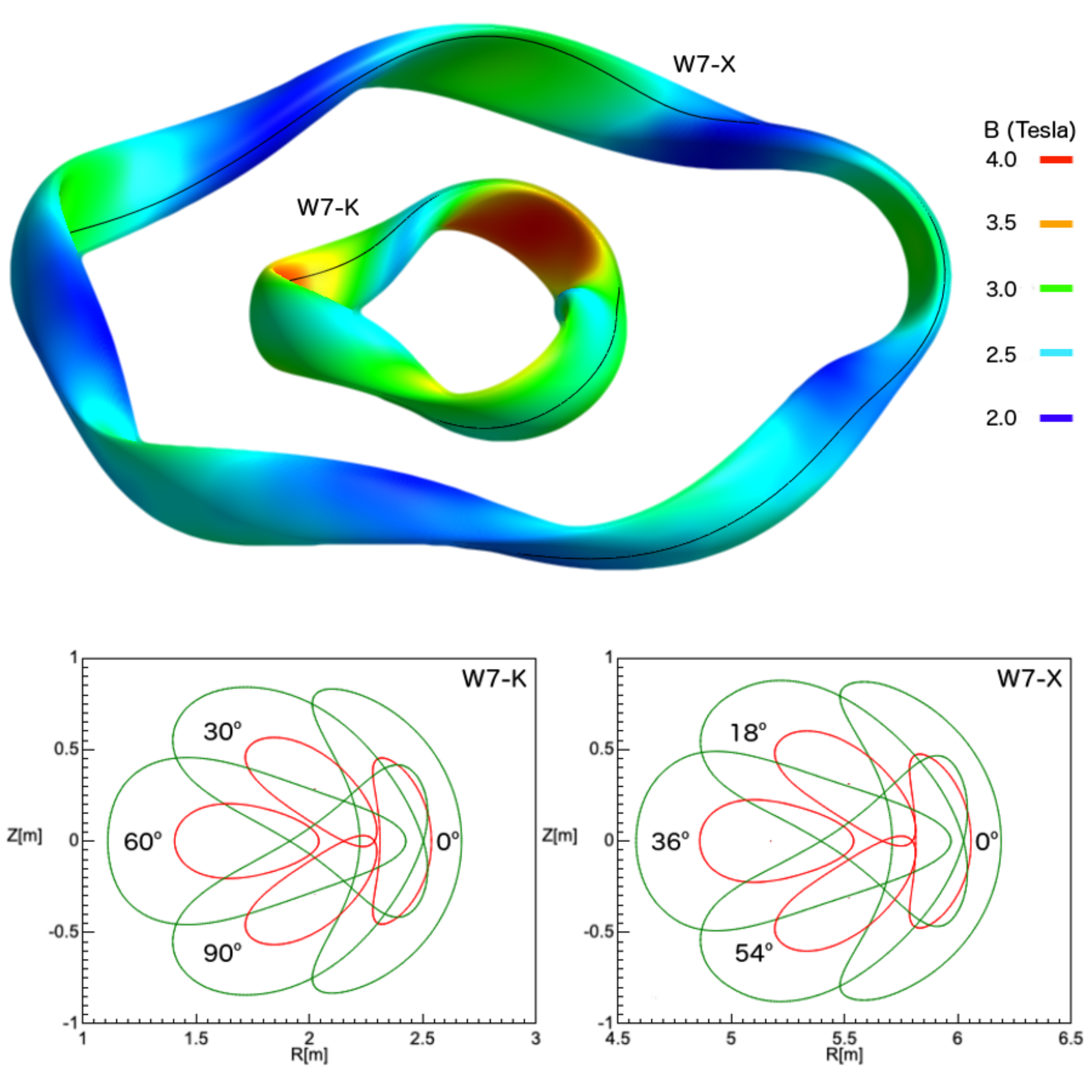}
    \caption{Comparing the W7-X high mirror configuration with W7-K. (Upper) Surfaces ($r/a=1$) showing $B$ in Tesla, with a single magnetic field line (solid black line) overlaid. (Lower) Cuts of magnetic surfaces at specified toroidal angles mapped onto the cylindrical coordinates $Z$ versus $R$. The cuts at $r/a=1$ are plotted in green while those at $r/a=0.5$ are in red. By construction the surface $r/a=1$ has the same shape for both configurations aside from a 5/3 rescaling of the toroidal angle.}
    \label{fig:W7K}
\end{figure}

\textit{Increasing the critical gradient.}-- Applying principles gleaned from the model equation (\ref{eqn:critgrad}) we devise two novel configurations.  The first is generated by modifying two geometric parameters of the high-mirror configuration of the Wendelstein 7-X stellarator to improve the critical gradient.  First, we directly control the effective radius of curvature, as compared to the minor radius, by reducing the aspect ratio of the configuration.  Specifically, we create a configuration whose outer flux surface shape is the same as the W7-X configuration, except for a rescaling of the major radius from $5.51$ m down to $2.00$ m.  This involves only modifying the $m = n = 0$ Fourier component of the radial component of surface shape specification in cylindrical cooordinates, {\em i.e.} $R_{00}$ in the input file to the equilibrium code VMEC \citep{Hirshman1983a}. Second, we reduce the field period number of the configuration from $5$ down to $3$. The field period reduction broadens the parallel connection length $L_{w}$, which we find stabilizes more extended ITG modes.  As Fig. \ref{fig:critgrad} shows, the resulting low sheared configuration, ``Wendelstein 7 Kompakt" (W7-K), has twice the ITG critical gradient of the high-mirror W7-X configuration. The primary reason is a near doubling of the drift curvature. Figure \ref{fig:W7K} shows the shape of the outermost flux surfaces for W7-X and W7-K, colored according to $B$, as well as surface cuts at evenly spaced toroidal angles. Furthermore, W7-K has values of $\epsilon_{\text{eff}}$ below $0.01$ out to the surface with $r^{2}/a^{2}=0.5$. This configuration thus demonstrates that the ITG mode threshold for a given stellarator configuration can be increased without spoiling neoclassical confinement. It also contains a reduction in the average value of $|\bnabla s|^{2}$, implying smaller linear growth rates for the ITG mode above marginality \citep{Xanthopoulos2014a}.  

\begin{figure}
    \centering
    \includegraphics[scale=.42]{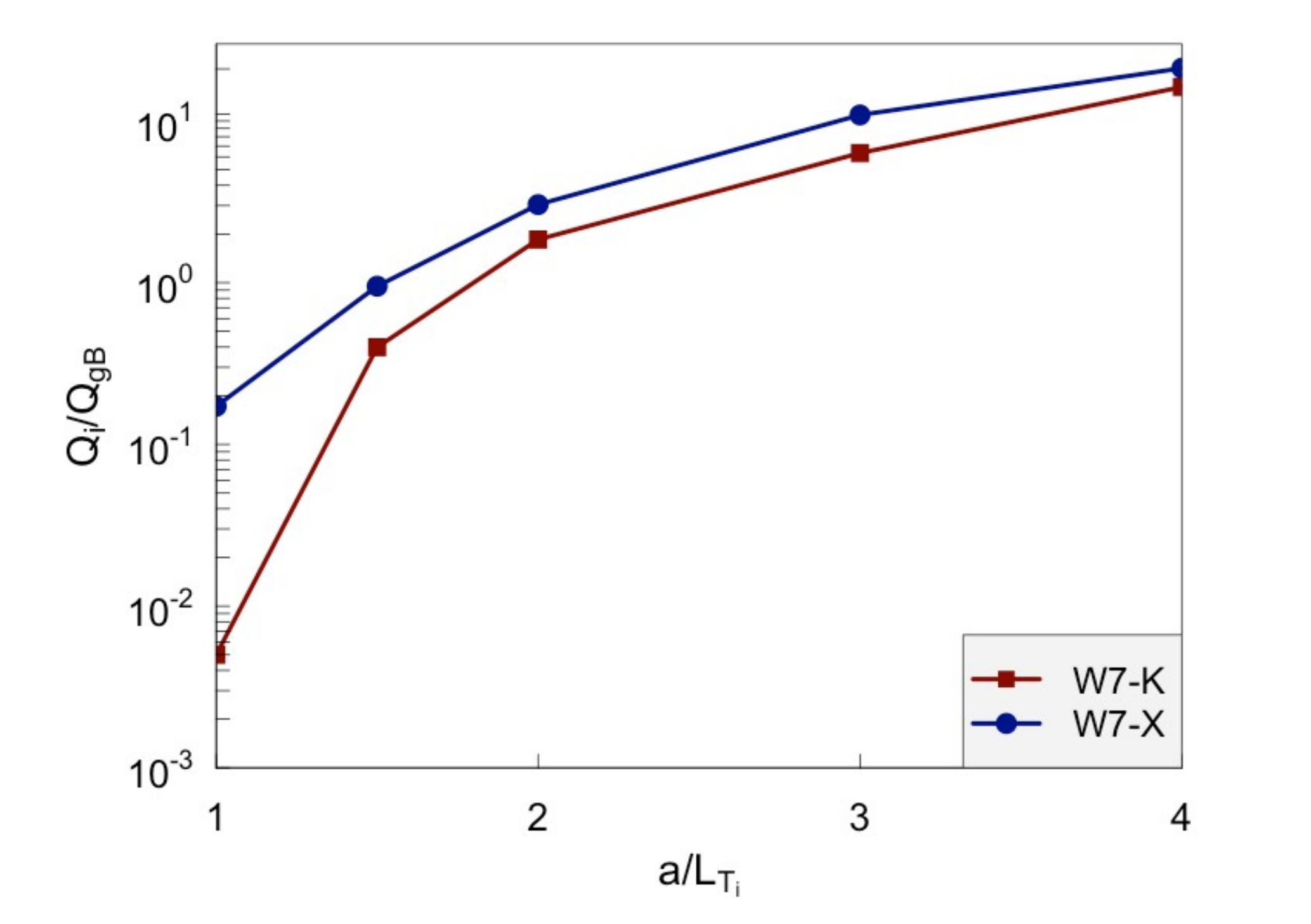}
    \caption{Nonlinear heat fluxes computed by the GENE code (adiabatic electrons, zero density gradient, $T_{e}=T_{i}$) for flux tube simulations at $\alpha=0$ (bean cross section, toroidal angle $0$) comparing the high mirror W7-X configuration to W7-K. The time-averaged heat steady-state heat flux $Q_{i}/Q_{GB}$ is plotted on a log-linear scale versus $a/L_{T,i}$, where $Q_{gB} = (\rho^{2}_{s}/a^{2}) c_{s} P_{i}$ is the gyro-Bohm heat flux, $c_{s}=\sqrt{T_{e}/m_{i}}$, $\rho_{s}=c_{s}/\Omega_{i}$, and $P_{i}=nT$.}
    \label{fig:ITG}
\end{figure}

The benefits of W7-K, however, are not limited to linear physics effects when considering ITG mode transport. We find two significant improvements in W7-K over W7-X via nonlinear GENE calculations of the time-averaged ion heat flux $ Q_{i}$, shown in Fig. \ref{fig:ITG}. Firstly, the nonlinear threshold for instability $a/L_{T,\text{crit}}$, inferred by extrapolation, is larger for W7-K, so the strategy of improving the linear critical gradient can lead to increases in its nonlinear counterpart. Secondly, the heat fluxes above the nonlinear threshold are consistently lower in the case of W7-K -- at a gradient $a/L_{T,i}=2$, the heat flux is nearly a factor of two smaller. Thus a reduction in transport as well as an increase in the nonlinear threshold has been simultaneously achieved. We attribute both improvements to the enhanced nonlinear zonal flow drive present in W7-K, which is outside the scope of the present discussion but will be explored in a future publication.

A second ``Nearly Axi-Symmetric'' (NAS) configuration  is generated by using the method described in \citet{Plunk2018}, yielding a weakly non-axisymmetric vacuum field that satisfies quasi-axisymmetry (QA) approximately, with an aspect ratio of $3$.  As shown in Fig.~\ref{fig:critgrad}(a), NAS enjoys the largest critical gradient of the QA configurations studied here.  This is due to the relatively large and uniform curvature on the outboard side of the device, in contrast to NCSX, whose shaping actually leads to relatively weak curvature, which is however compensated by strong negative shear.

\textit{Conclusions.}-- A coarse-graining gyrokinetic model has been implemented to estimate the critical gradient for the ITG mode in a large variety of stellarator configurations.  The comparison against calculations with the GENE code confirms the robustness of this approach, even for configurations whose geometry is not ideally represented by the ``smoothed'' drift
curvature. 

We note that the coarse-graining method becomes less accurate when scale separation is absent, as in the case of QPS [Fig. 2(d)], or when global shear becomes extremely
small. The latter limit is not approached in the cases analyzed here, as we have not taken flux tubes at inner radii $r^{2}/a^{2} < 0.5$.  Our investigation of the low-shear regime, to be published later, confirms that the critical gradient is set by Floquet-like, extended modes \citep{Bhattacharjee1983,Zocco2018,Faber2018} in such cases. 

The main benefit of this novel technique is twofold: i) A swift prediction of the critical gradient without the need for tedious gyrokinetic numerical calculations, and ii) the potential to construct new configurations that combine neoclassical optimization with a high ITG critical gradient. Two such configurations found here are a compact version of the Wendelstein 7-X, and a low aspect ratio quasi-axisymmetric configuration. Our results suggest that, using the ITG critical gradient as a figure of merit, a smaller aspect ratio can be beneficial, but there is ample room among other degrees of freedom for improvement in future designs of fusion reactors of the stellarator type.

The authors thank P. Helander, A. Zocco, and M. Landreman for helpful conversations. This work was supported by a grant from the Simons Foundation (No. 560651, G. T. R.-C.). Computing resources at the Cobra cluster at IPP Garching and the Marconi Cluster were used to perform the simulations.  This work has been carried out within the framework of the EUROfusion Consortium, funded by the European Union via the Euratom Research and Training Programme (Grant Agreement No 101052200 — EUROfusion). Views and opinions expressed are however those of the author(s) only and do not necessarily reflect those of the European Union or the European Commission. Neither the European Union nor the European Commission can be held responsible for them.

\bibliography{library}


\end{document}